# Sequence correlations shape protein promiscuity

David B. Lukatsky[1*], Ariel Afek[1], and Eugene I. Shakhnovich[2*]
[1]*Department of Chemistry, Ben-Gurion University of the Negev, Beer-Sheva 84105 Israel*
[2]*Department of Chemistry and Chemical Biology, Harvard University*

**Abstract**

We predict analytically that diagonal correlations of amino acid positions within protein sequences statistically enhance protein propensity for nonspecific binding. We use the term 'promiscuity' to describe such nonspecific binding. Diagonal correlations represent statistically significant repeats of sequence patterns where amino acids of the same type are clustered together. The predicted effect is qualitatively robust with respect to the form of the microscopic interaction potentials and the average amino acid composition. Our analytical results provide an explanation for the enhanced diagonal correlations observed in hubs of eukaryotic organismal proteomes [J. Mol. Biol. **409**, 439 (2011)]. We suggest experiments that will allow direct testing of the predicted effect.

## I. INTRODUCTION

Recent experimental evidences that proteins within a cell maintain a high degree of nonspecificity have challenged the understanding of molecular mechanisms providing the specificity of protein-protein binding [1-3]. Such nonspecific binding is often termed protein 'promiscuity'. Numerous organismal-scale measurements of binary protein-protein interactions (PPI) suggest that organismal proteomes possess a higher degree of nonspecific binding [4-10]. The dominant amount of such experimental PPI data come from the high-throughput yeast two-hybrid (Y2H) [7, 11, 12] and affinity purification followed by mass spectrometry (AP/MS) [13-15]. Although these experiments do not provide dynamical and functional properties of interactions [7], they do provide a snapshot of binary, physical interactions that might be functionally important in a living cell. These experiments show that a small fraction of proteins (termed 'hubs') are physically interacting with tens and even hundreds of partners. It appears that protein promiscuity is a selectable trait enabling proteins to adopt more efficiently to changing conditions or emerging needs within a living cell [1].

---
* Corresponding author: E. I. S.: shakhnovich@chemistry.harvard.edu
             D. B. L.: lukatsky@bgu.ac.il

The key question is what makes a protein promiscuous, *i.e.*, prone to nonspecific interactions? Are there generic sequence signatures of promiscuity? In this paper we predict one such generic signature. We show analytically that protein sequences with enhanced correlations of sequence positions of amino acids of the same type generally represent more promiscuous sequences. We use the term 'diagonal' to describe such correlations. Intuitively, sequence 'correlations' mean statistically significant repeats of sequence patterns. Our findings suggest that symmetry properties and strength of sequence correlations are important factors that control global connectivity properties of PPI networks.

Using a bioinformatics analysis, we have recently shown that hubs of eukaryotic (*e.g.*, yeast and human) proteomes possess a higher level of diagonal correlations compared to non-hubs [16]. In particular, we have shown that in human PPI network, His, Phe, Ile, Pro, Gly, and Tyr, exhibit significantly stronger diagonal correlations in hubs than in non-hubs [16]. Here, we develop an analytical theory that explains why this might be the case. Since many hub proteins detected in Y2H and AP/MS experiments are confirmed to be functionally multi-specific *in vivo*, our theoretical predictions suggest that enhanced nonspecific binding and functional multi-specificity might be tightly linked. We suggest that a significant fraction of functionally multi-specific proteins might be inherently highly promiscuous.

This article is organized as follows. First, we introduce a precise, statistical measure of promiscuity. Second, we present a simple model that describes protein-protein binding. This model uses a two-letter alphabet, linear protein sequences, and it is analytically solvable in the Gaussian approximation. We develop a stochastic procedure allowing us to design protein sequences with a controlled symmetry and strength of sequence correlations. We analyze statistical, interaction properties of such sequences. Third, we qualitatively describe possible implications of our results for protein folding. We conclude by proposing experiments that will allow direct testing of the predicted effect.

## II. RESULTS AND DISCUSSION

### A. Statistical measure for interaction promiscuity

We begin by introducing a statistical measure of promiscuity for a given protein, *A*. Such measure can be defined as the probability distribution of the interaction energies, $P(E_A)$, of this protein with a set of target proteins, where $E_A$ is the interaction energy between protein *A* and a protein from the target set. Now we can compare the promiscuity of two proteins *A* and *B* interacting with the same target set, assuming that the corresponding average interaction energies are the same, $\langle E_A \rangle = \langle E_B \rangle$. This latter target set is not supposed to be optimized in any way for stronger binding with either protein *A* or protein *B*. If the dispersion, $\sigma_A$, of $P(E_A)$ is greater than the dispersion, $\sigma_B$, of $P(E_B)$, then protein *A* is statistically more promiscuous than protein *B*. This is because if $\sigma_A > \sigma_B$, the mean free energies of binding obey the inequality, $F_A < F_B$ [17]. In particular, we have recently shown analytically that if $P(E_A)$ and $P(E_B)$ are Gaussian, then the mean free energy difference always satisfies, $F_A - F_B = -(\sigma_A^2 - \sigma_B^2)/2k_B T$ [17]. The assumption that $\langle E_A \rangle = \langle E_B \rangle$ corresponds to the constraint that the sequences of two proteins *A* and *B* have the same average amino acid composition (see below). The latter constraint is necessary for a fair comparison of promiscuities, since the differences in the average amino acid composition would produce a trivial shift of the average interaction energies. The predicted effect is induced exclusively by sequence correlations and goes beyond the mean-field. The above argument holds true if instead of two proteins *A* and *B*, we consider two sets of proteins $(A_1,...,A_M)$ and $(B_1,...,B_M)$ interacting with the same set of random binders, as above, and characterized by analogous distributions of the binding energies, $P(E_A)$ and $P(E_B)$. Our objective is to compute the properties of $P(E)$ for model proteins with controllable strength and symmetry of sequence correlations, interacting with a set of random binders, reflecting the statistics of interaction strengths in crowded cellular environments.

**B. Analytical model for 'random' and 'designed' interacting sequences**

We introduce now a simple model for 'random' and 'designed' (or 'correlated') protein-like, linear sequences, Fig. 1. Despite its one-dimensional origin, the model is not exactly solvable because of the generally long-range nature of the potentials that we use below. For simplicity we use a minimalistic sequence alphabet with only two types of amino acids. Random sequence is obtained by distributing $N_p$ and $N_h$ amino acids of types P and H, respectively, at random within a linear sequence of the total length $L = N_p + N_h$. Our simplistic approach therefore does not take into account the folding of the sequence. The average linear fraction of P and H amino acids is thus fixed and given by $\phi_{p,0} = N_p/L$ and $\phi_{h,0} = N_h/L$, respectively. The notion of P and H types stands here just in order to distinguish between two different amino acid types, and it does not constrain our conclusions to just hydrophobic and polar types. Our conclusions hold qualitatively true for any number of amino acid types.

After each random sequence is generated, amino acid identities are fixed and not allowed to change their positions. A correlated sequence is obtained using the following stochastic procedure. First, we generate a random sequence as described above. Second, we allow amino acids to anneal at a given 'design' temperature, $T_d$. We note that our notion of the 'designed' sequences stands to describe the existence of positional correlations of amino acids within linear sequences and not the folding. We thus impose that amino acids within the sequence under the design procedure interact through the pair-wise additive design potential, $U_{\alpha\beta}(x)$. The intra-sequence interaction energy for any given amino acid realization:

$$E_{\text{intra}} = \frac{1}{2}\int \phi_p(x) U_{pp}(x-x')\phi_p(x')\,dx\,dx'$$
$$+ \frac{1}{2}\int \phi_h(x) U_{hh}(x-x')\phi_h(x')\,dx\,dx' \qquad \text{Eq. (1)}$$
$$+ \int \phi_h(x) U_{hp}(x-x')\phi_p(x')\,dx\,dx'$$

where $\phi_p(x)$ and $\phi_h(x)$ are the local, linear fraction densities of P and H amino acids, respectively. The average composition of P and H amino acids is fixed by the values, $\phi_{p,0}$ and $\phi_{h,0}$, respectively, and we impose that the total fraction of amino acids at each sequence position, $x$, is unity, $\phi_p(x) + \phi_h(x) = 1$. Here $U_{pp}(x)$, $U_{hh}(x)$, and $U_{hp}(x)$ is the interaction potential between PP, HH, and HP amino acid pairs, respectively. We also note that $\phi_p(x)$ can be represented in the form: $\phi_p(x) = \phi_{p,0} + \delta\phi_p(x)$, where $\delta\phi_p(x)$ is the deviation of the local density of P-type amino acids from its average value, and analogously, $\phi_h = \phi_{h,0} + \delta\phi_h(x)$. The only two assumptions about the interaction potentials, $U_{\alpha\beta}(x)$, used in the sequence design procedure are that they are pair-wise additive and have a finite range of action.

We note that below, we also use the Monte-Carlo (MC) stochastic design procedure that performs actual amino acid swaps with the Metropolis criterion for the energy change [18].

Our next step is to analyze the probability distribution $P(E)$ of the interaction energy, $E$, between random and correlated sequences. Every pair of interacting sequences thus consists of one random and one correlated sequence superimposed in a parallel configuration, thus the problem is a quasi one-dimensional one. We show below that enhanced correlations between amino acids of the same type lead to the broadening of the distribution, $P(E)$. The resulting binding free energy computed from the broader distribution will be lower than the binding free energy computed from a narrower $P(E)$ [17]. The latter property implies that such correlated sequences will be more promiscuous, *i.e.* statistically prone to a stronger binding with an arbitrary sequence. We use an ensemble of entirely random sequences as a proxy for an ensemble of arbitrary protein sequences. The interaction energy between the random and correlated sequences:

$$E = \int v_p(\rho) V_{pp}(\rho - \rho') \phi_p(\rho') d\rho d\rho' + \int v_h(\rho) V_{hh}(\rho - \rho') \phi_h(\rho') d\rho d\rho'$$
$$+ \int v_h(\rho) V_{hp}(\rho - \rho') \phi_p(\rho') d\rho d\rho' + \int v_p(\rho) V_{ph}(\rho - \rho') \phi_h(\rho') d\rho d\rho'$$

Eq. (2)

where $v_p(\rho)$ and $v_h(\rho)$ are the local, linear fraction densities of P and H amino acids, respectively, within the random sequence, and $\rho$ is the inter-sequence distance. Here again, $v_p(\rho) + v_h(\rho) = 1$, and $v_p(\rho) = \phi_{p,0} + \delta v_p(\rho)$, with $\delta v_p(\rho)$ being the deviation of the P-type amino acid density from its average value, and analogously for $v_h(\rho) = \phi_{h,0} + \delta v_h(\rho)$. We thus assume that the average amino acid composition is the same for random and correlated sequences. We emphasize that the inter-sequence interaction potentials, $V_{pp}(\rho)$, $V_{hh}(\rho)$, and $V_{hp}(\rho) = V_{ph}(\rho)$, need not be identical to the potentials $U_{pp}(x)$, $U_{hh}(x)$, and $U_{hp}(x)$ used in the sequence design procedure to generate sequences with a controlled symmetry and strength of correlations. We now describe in details the effect of the potentials $V_{\alpha\beta}(\rho)$ and $U_{\alpha\beta}(x)$ on the properties of $P(E)$.

The probability distribution for the interaction energies between the random and correlated sequences, $P(E)$, is characterized by its mean, $\langle E \rangle$, and by the variance, $\sigma^2$. In particular, in order to compute $P(E)$, we generate the ensemble of interacting sequence pairs, where each pair consists of a random sequence and 'designed' (*i.e.* correlated) sequence. The mean, $\langle E \rangle$, is independent on the design potential, $U_{\alpha\beta}(x)$, and therefore all the different distributions $P(E)$ obtained at different values of the design temperature, $T_d$, will have exactly the same mean. The variance of $P(E)$, $\sigma^2 = \left\langle (\delta E_2)^2 \right\rangle$, where the only relevant term for the averaging is quadratic in the sequence density fluctuations:

$$\delta E_2 = \int \delta v_p(\rho) V(\rho - \rho') \delta \phi_p(\rho') d\rho d\rho', \qquad \text{Eq. (3)}$$

where $V(\rho) = V_{pp}(\rho) + V_{hh}(\rho) - 2V_{hp}(\rho)$, and $\hat{V}(k) = \int V(\rho) e^{-ik\rho} d\rho$. The averaging in $\left\langle (\delta E_2)^2 \right\rangle$ is performed using the Boltzmann probability distribution function for the sequence density fluctuations of the correlated (*i.e.* designed) sequences:

$$P_d[\delta\phi_p(x)] = C_1 \exp\left[-\int \frac{\delta\phi_p^2(x)}{2\phi_{p,0}\phi_{h,0}} dx\right] \exp(-E_{\text{intra}}/k_B T_d). \qquad \text{Eq. (4)}$$

The first exponential term in Eq. (4) is the entropic contribution [19] due to the sequence density fluctuations of the 'designed' sequences, and the second exponential term represents the strength of the correlations within the 'designed' sequences. The corresponding probability distribution for the density fluctuations of the random sequences contains only the entropic contribution:

$$P_r[\delta v_p(\rho)] = C_2 \exp\left[-\int \frac{\delta v_p^2(\rho)}{2\phi_{p,0}\phi_{h,0}} d\rho\right]. \qquad \text{Eq. (5)}$$

The constants $C_1$ and $C_2$ in Eqs. (4) and (5), are found from the normalization constrains applied on the probability distributions. The averaging leads to the following result:

$$\sigma^2 = 4L\phi_{p,0}\phi_{h,0} \int \frac{dk}{2\pi} \left|\hat{V}(k)\right|^2 \frac{1}{1/\phi_{p,0}\phi_{h,0} + \hat{U}(k)/k_B T_d}, \qquad \text{Eq. (6)}$$

where $\hat{U}(k) = \int U(x) e^{ikx} dx$, and $U(x) = U_{pp}(x) + U_{hh}(x) - 2U_{ph}(x)$. The larger $\sigma$ (and thus the broader the distribution $P(E)$ of the interaction energies between the correlated and random sequences), the more promiscuous are the correlated sequences. We note that our model is only solvable analytically in the Gaussian approximation, and not exactly solvable, unlike the one-dimensional Ising model, due to the generally long-range nature of the intra-sequence ('design') potential, $U(x)$, and the inter-sequence potential, $V(\rho)$. We also note the existence of the singularity in Eq. (6) at sufficiently large and negative values of the 'design' potential, $U(x)$, when the Gaussian fluctuation model breaks down.

The analysis of Eq. (6) leads to the two key conclusions. First, the more negative is the 'design' potential, $U(x)$, the larger is $\sigma$. Taking into account the definition of $U = U_{pp} + U_{hh} - 2U_{ph}$, one concludes that in order to increase $\sigma$ one needs to design the sequences with the enhanced correlations in the positions between the residues of similar types. This means that correlated sequences where amino acids of the same type are clustered together will be the more promiscuous ones. Second, such correlated sequences will interact statistically stronger (than non-correlated sequences would do) with any arbitrary sequences independently on the sign of the inter-residue interaction potential, $V = V_{pp} + V_{hh} - 2V_{ph}$. Third, if the design potential is overall positive, $U > 0$, designed sequences will be even less promiscuous than random sequences. We emphasize that the predicted effects are generic and qualitatively independent on the specific form and even sign of the microscopic interaction potentials, $V_{\alpha\beta}$, and on the average amino acid composition of the sequences.

We note that the predicted effect gets even stronger when both interacting sequences are 'designed' (*i.e.* correlated). In the latter case the variance, $\sigma_{d,d}$, of the corresponding $P(E)$ is a straightforward generalization of Eq. (6):

$$\sigma_{d,d}^2 = 4L \int \frac{dk}{2\pi} \left|\hat{V}(k)\right|^2 \frac{1}{(1/\phi_{p,0}\phi_{h,0} + \hat{U}_1(k)/k_BT_{d1})(1/\phi_{p,0}\phi_{h,0} + \hat{U}_2(k)/k_BT_{d2})}, \qquad \text{Eq. (7)}$$

where $U_1(x)$ and $U_2(x)$ are defined analogously to $U(x)$ for each of the interacting sequences; and $T_{d1}$ and $T_{d2}$ are the design temperatures for the first and second sequence, respectively. If both 'design' potentials, $U_1(x)$ and $U_2(x)$, are overall negative, then $\sigma_{d,d} > \sigma$, and thus in the latter case the sequences will be statistically more promiscuous than in the case when only one of the interacting sequences is 'designed' (Eq. (6)). We stress that the interacting sequences are designed independently and not optimized in any way towards stronger binding. Therefore the observed effect of statistically enhanced binding corresponds to nonspecific (promiscuous) binding. We note that in both cases, Eq. (6) and Eq. (7), the greatest $\sigma$ and $\sigma_{d,d}$, are achieved when the average amino acid composition of sequences is uniform, $\phi_{p,0} = \phi_{h,0} = 1/2$.

In order to verify our theoretical predictions, we first perform the standard MC annealing procedure [18] to design correlated sequences. We begin with generating a random sequence starting with a given amino acid composition. We next perform the MC stochastic design procedure, where amino acids within the sequence are allowed to exchange their positions, and each sequence configuration has the Boltzmann weight, $\sim \exp(-E_{\text{intra}}/k_BT_d)$, where $E_{\text{intra}}$ is the internal energy of the sequence in a given configuration given by Eq. (1). The MC design procedure is stopped after a certain number of MC moves, and the resulting annealed configuration is accepted as the final, designed configuration for a given sequence.

The lower $T_d$ is, the stronger are the correlations within the sequences. Intuitively, stronger correlations correspond to repetitive sequence patterns with a longer correlation length. The properties of the correlated patterns depend critically on the sign of the interaction potentials $U_{\alpha\beta}(x)$ used in the design procedure. If the effective design potential $U = U_{pp} + U_{hh} - 2U_{hp}$ is overall negative (this corresponds to the attraction between the amino acids of similar types), the correlated patterns will have the form of repetitive residues of the same type, for example: HHHHPPPPHHHPPP… If however, the potential $U = U_{pp} + U_{hh} - 2U_{hp}$ is overall positive, the correlated patterns will have the form of the alternating H and P residues, for example: HPHPHPHPHPHP… To characterize the correlation properties of the sequences quantitatively, we introduce the normalized correlation function:

$$\eta_{\alpha\beta}(x) = g_{\alpha\beta}(x) / \langle g^r_{\alpha\beta}(x) \rangle_r, \qquad \text{Eq. (8)}$$

where $g_{\alpha\beta}(x)$ is proportional to the probability to find a residue of the type $\alpha$ separated by the distance $x$ from a residue of the type $\beta$, and $g^r_{\alpha\beta}(x)$ is the corresponding probability for the randomized sequence, and $\langle g^r_{\alpha\beta}(x) \rangle_r$ corresponds to the averaging with respect to different realizations of randomized sequences. The computed correlation functions are represented in Fig. 2 at the value of $k_B T_d = 1$ (in the units of $k_B T$). For the entirely uncorrelated (random) sequences, all the matrix elements of $\eta_{\alpha\beta}(x)$ are equal to unity, Fig. 2. The clustering of amino acids of similar types corresponds to $\eta_{\alpha\alpha}(x) > 1$, Fig. 2.

The next step is to compute numerically the properties of the probability distribution, $P(E)$, of the interaction energies, $E$, between random and designed sequences (*i.e.* each interacting pair consists of a random and designed sequences). The results of these calculations are shown in Fig. 3. We computed $P(E)$ at different values of $T_d$ and we represented the results as a ratio between the dispersion of $P(E)$, $\sigma = \sigma_{d,r}$ and the dispersion of the corresponding probability distribution where *both* sequences are entirely random, $\sigma_{r,r}$ (the latter corresponds to the case of a vanishing design potential, $U_{\alpha\beta} = 0$). We used here the inter-residue interaction potential, $V_{pp} = V_{hh} = -1$, and $V_{hp} = 1$, and we assumed that the nearest neighbor and the next-nearest neighbor amino acids can interact between the two sequences. The analytical result computed from Eq. (6) is also plotted in Fig. 3. As expected, the Gaussian fluctuation model becomes accurate at small values of the ratio, $U(a)/k_B T_d \ll 1$, where $a$ is the potential range. The insert of Fig. 3 shows the computed $P(E)$ in the case of designed-random and random-random sequence pairs, respectively [16]. The key conclusion here is that in accordance with the analytical predictions, the dispersion of $P(E)$ is larger for the sequences designed with the overall negative $U$, as compared to the dispersion of $P(E)$ in the case where both interacting sequences are entirely random, $\sigma_{d,r} > \sigma_{r,r}$. The effect is the opposite for overall positive $U$, $\sigma_{d,r} < \sigma_{r,r}$. We stress that these conclusions are qualitatively insensitive to the sign of the inter-residue interaction potential, $V(x)$. We emphasize also that the interaction energy mean-value, $\langle E \rangle$, is identical in all cases, it depends on the average amino acid composition of sequences, and it is insensitive to the correlation properties of interacting sequences.

## C. Implications of sequence correlations for protein folding

A key limitation of our theoretical analysis is the fact that the presented simplified model does not explicitly take into account protein folding, and therefore, underestimates the effect of longer-range sequence correlations induced by the presence of a protein chain. Taking protein folding into account should provide additional insights into the effect of long-range sequence correlations on protein promiscuity and structural disorder. Elucidation of the latter issue is the subject of our future work.

Using a bioinformatics analysis, we have recently shown that enhanced diagonal sequence correlations are strongly overrepresented in structurally disordered proteins, as compared to structurally ordered proteins (such as all-alpha and all-beta proteins) [16]. In particular, we have observed that in a set of experimentally known disordered proteins, diagonal correlations are significantly enhanced for Gly, Tyr, Arg, Trp, Ser, Glu, Pro, Asp, Gln, Ala, Lys, and Thr [16].

Our simplified analysis suggests that such enhanced diagonal correlations generically widen the energy spectrum of nonspecific states within the proteins, which leads to the lowering of the energy for disordered conformations. We use the term 'intra-protein promiscuity' to describe increased probability for thermodynamically allowed, nonnative conformational states. Specifically, it was shown [20-23] that statistics of energies of misfolded conformations which are structurally dissimilar to native state, obeys the Random Energy Model [24], so that the energy gap between the native state and the lowest energy misfolded conformation can be estimated as:

$$\Delta = E_{nat} - \langle E \rangle - \frac{\sigma}{\sqrt{2\ln\gamma}} \qquad \text{Eq. (9)}$$

where $E_{nat}$ is effective free energy of native state (which incorporates energetic and entropic solvent effects, as well as the entropy of small motions around the native conformation), $\sigma$ is the standard deviation of the energies of misfolded conformations, and $\gamma$ is effective number of conformations per amino acid [25]. The physical intuition behind Eq. (9) is that the total energy of a protein in any conformation is a sum of energies of many interacting fragments leading to the Gaussian distribution of energies of protein conformations. To that end, the one-dimensional model developed in this paper applies to the interaction of fragments, providing a corrected value of $\sigma$. In particular, diagonal correlations increase the variance of energies of misfolded conformations, decreasing the energy gap, $\Delta$, which leads to a higher likelihood that the protein chain will end up in a misfolded conformation and a slower folding rate to the native state [26, 27]. While these considerations are suggestive, a straightforward analysis based on the replica theory of protein-like heteropolymers [28] is required to quantitatively assess the impact of diagonal correlations on stability of unique folded states of proteins. Such analysis is forthcoming.

## III. CONCLUSIONS

In summary, we predict here analytically that sequence correlations statistically shape the energy spectrum of nonspecific inter-protein binding (we term such nonspecific binding 'inter-protein promiscuity'). In particular, correlated sequences with enhanced diagonal correlations of the type HHHHHPPPPPPHHHHHPPPP, where amino acids of the same type are clustered together will bind statistically stronger to an arbitrary target sequence set, compared to either random sequences, or correlated sequences of the type HPHPHPHPHPHPHPHPHP. Sequences possessing the latter symmetry of correlations will constitute the least promiscuous sequences. In general, the longer is the length scale of homo-oligomer repeats, the wider the inter-protein interaction energy spectrum, and the more promiscuous are the sequences. This effect is qualitatively robust with respect to the specific form and even sign of the microscopic inter-sequence interaction potential, and it is controlled by the length scale and symmetry of sequence correlations. Despite the one-dimensional nature of our model, its results are directly applicable to protein-protein interaction networks since the most recent, whole-organism experimental and bioinformatics data suggest that 15-40% of all protein-protein interactions are mediated by linear sequence motifs, and not by large protein surfaces [29]. Our analytical predictions provide an explanation for the enhanced diagonal correlations observed in hubs of human and yeast PPI networks [16]. Our key objective for the future theoretical analysis is to take into account the effect of protein folding on inter-protein interaction properties.

There are several possible strategies to test our predictions. The direct experimental test would utilize a protein chip [30] or microfluidic protein chip [31] technology. The target protein data set would be attached to the chip surface. The test proteins or peptides would be synthesized with a varying strength and symmetry of sequence correlations but keeping the average amino acid composition fixed. Titration experiments should allow measuring directly the binding affinity [30, 31] as a function of sequence correlation properties. We expect that protein sequences with enhanced diagonal correlations will generically represent more promiscuous sequences. Another possibility is to use a recent genome-wide protein over-expression analysis [32]. Since the over-expression of highly promiscuous proteins should presumably be toxic to a cell, the correlation analysis of such toxic proteins (hundreds of them are known [32]) will show whether the predicted effect plays a significant role in a living cell [33].

**ACKNOWLEDGEMENTS**

D. B. L. acknowledges the financial support from the Israel Science Foundation (ISF) grant 1014/09. A. A. is a recipient of the Lewiner graduate fellowship.

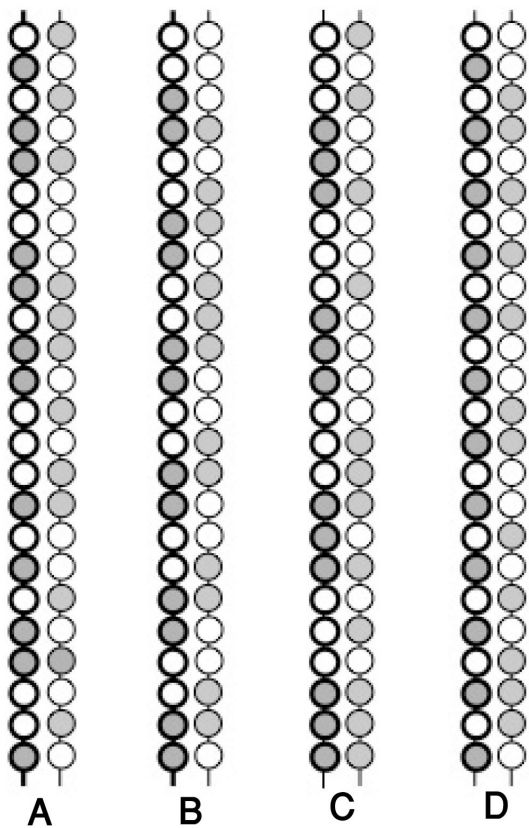

**Figure 1:** Cartoon illustrating different types of sequence correlations. Right protein in each interacting pair represents a 'random binder' (randomly distributed amino acids of two types). Left protein represents: **A.** Random sequence (no correlations). **B. and C.** Correlated sequence (amino acids of the same type have a tendency to cluster; we term such correlations 'diagonal'). Such sequences are predicted to possess higher propensity for nonspecific binding. Sequence correlations in **(C)** have longer correlation length and such sequence is statistically more promiscuous than **(B)**. **D.** Correlated sequence where amino acids of the same type are alternating along the sequence. Such sequence represents the least promiscuous sequence. The symmetry of sequence correlations in **(D)** is fundamentally different compared to **(B)** and **(C)**. In the analytical calculations we consider many statistical realizations of sequences for each type of correlation symmetry.

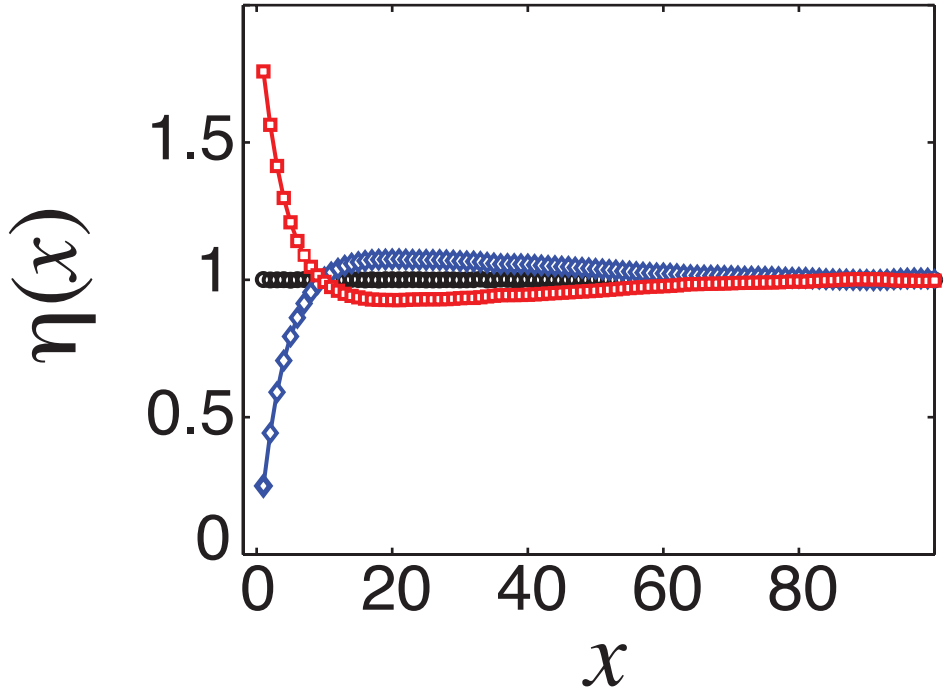

**Figure 2:** Computed sequence correlation functions for the 'designed' sequences at $T_d = 1$ (in the units of $k_B T$), $\eta_{pp}(x) = \eta_{hh}(x)$ (red squares), $\eta_{ph}(x) = \eta_{hp}(x)$ (blue diamonds); and for the random sequences, (black circles). All the matrix elements of $\eta_{\alpha\beta}(x)$ are the same for the random sequences. The design potential was chosen to be $U_{pp} = U_{hh} = -1$, and $U_{hp} = 1$, and we assumed that only the nearest-neighbor residues can interact. The sequence length was chosen to be 200 amino acids, and we generated 5000 different sequences in each calculation. The plotted $\eta_{\alpha\beta}(x)$ represent the average over the entire set of the designed sequences. The uniform amino acid composition was adopted: 50% P and 50% H amino acids in each sequence. The error bars are smaller than the symbol size.

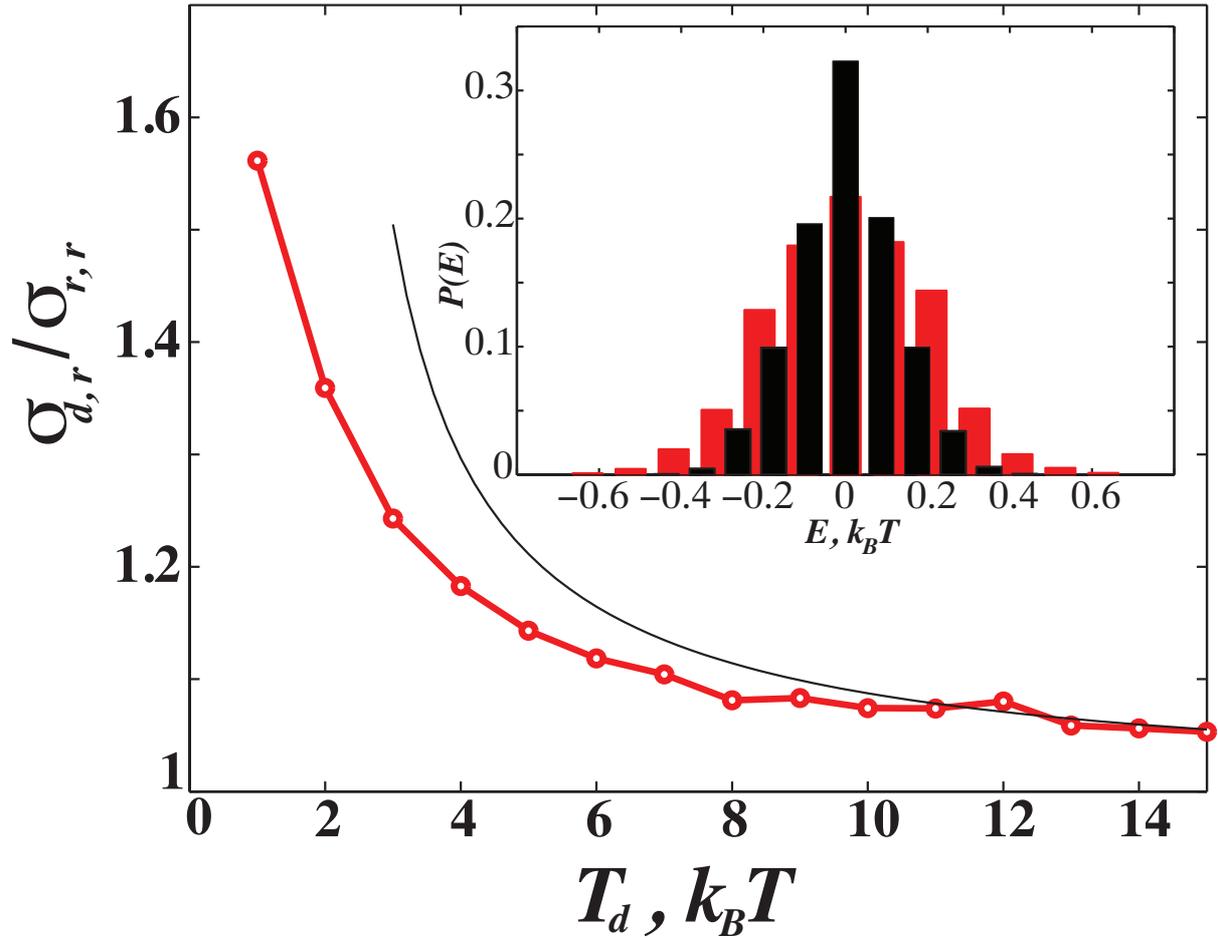

**Figure 3:** Computed ratio between the dispersions of the $P(E)$ for the interaction energies of the designed-random, $\sigma = \sigma_{d,r}$, and random-random, $\sigma = \sigma_{r,r}$, sequence pairs at different values of the design temperature, $T_d$ (circles). The error bars are smaller than the symbol size. The uniform amino acid composition was adopted: 50% P and 50% H residues in each sequence. Thin curve represents the corresponding analytical result, Eq. (6). Insert (adopted from Ref. [16]): Computed probability distribution function, $P(E)$, for the interaction energies between the pairs of two random sequences (black), and pairs consisting each of a random and a designed sequences, where the designed sequences were generated at $T_d = 1$ (in the units of $k_B T$) (red). The energy $E$ is normalized per one amino acid.